\documentclass[a4paper]{article}
\usepackage{Odyssey2020}
\usepackage{epsfig,amssymb,amsmath,lipsum,hhline,array,hyperref,algorithm2e,bm,float}
\ninept

\newcolumntype{x}[1]{>{\centering\arraybackslash\hspace{0pt}}p{#1}}
\DeclareMathOperator{\softmax}{softmax}

\title{DropClass and DropAdapt:\\ Dropping classes for deep speaker representation learning}

\makeatletter
\def\name#1{\gdef\@name{#1\\}}
\makeatother
\name{\em Chau Luu, Peter Bell, Steve Renals\thanks{Supported by an EPSRC iCASE studentship collaboration with the BBC.}}
\address{Centre for Speech Technology Research\\ University of Edinburgh, UK  \\
{\small \tt \{chau.luu, peter.bell, s.renals\}@ed.ac.uk} }
\begin{document}
\maketitle

\begin{abstract}
Many recent works on deep speaker embeddings train their feature extraction networks on large classification tasks, distinguishing between all speakers in a training set. Empirically, this has been shown to produce speaker-discriminative embeddings, even for unseen speakers. However, it is not clear that this is the optimal means of training embeddings that generalize well. This work proposes two approaches to learning embeddings, based on the notion of dropping classes during training.  We demonstrate that both approaches can yield performance gains in speaker verification tasks. The first proposed method, DropClass, works via periodically dropping a random subset of classes from the training data and the output layer throughout training, resulting in a feature extractor trained on many different classification tasks. Combined with an additive angular margin loss, this method can yield a 7.9\% relative improvement in equal error rate (EER) over a strong baseline on VoxCeleb. The second proposed method, DropAdapt, is a means of adapting a trained model to a set of enrolment speakers in an unsupervised manner. This is performed by fine-tuning a model on only those classes which produce high probability predictions when the enrolment speakers are used as input, again also dropping the relevant rows from the output layer. This method yields a large 13.2\% relative improvement in EER on VoxCeleb. The code for this paper has been made publicly available.
\end{abstract}

\section{Introduction}
% This template can be found on the conference website. Please use
% either one of the template files found on this website when preparing your
% submission. If there are special questions or wishes regarding
% paper preparation and submission for Odyssey 2020, correspondence
% should be addressed to \mbox{$<$info@odyssey2020.org$>$.}

% \begin{itemize}
%     \item Representation learning, discriminability. Large num classes
%     \item Comparison to multitask, multi objective training
%     \item Comparison to meta learning literature: training a good feature extractor, classification head is just acting upon a good universal feature extractor. MAML Loop comparison.
%     \item Comparison to dropout
%     \item Comparison to label smoothing techniques
% \end{itemize}

Deep speaker embeddings have become the state of the art technique in learning speaker representations \cite{Snyder,Snyder2018}, outperforming the historically successful i-vector technique \cite{Dehak2011}. These speaker representations are crucial for many tasks related to speaker recognition, such as speaker verification, identification and diarization \cite{Sell2018,Diez2019,Villalba2020}.

The networks used to generate speaker embeddings, such as the popular x-vector architecture \cite{Snyder2018}, are typically trained on a speaker classification task, taking as input the acoustic features of an utterance and predicting which training set speaker produced the input utterance. By taking one of the upper layers of this network as an embedding, a fixed dimensional vector can be extracted for any given input utterance. This vector, due to the training objective that the network was given, is speaker-discriminative. Crucially, it has been found that these embeddings can be used to discriminate between speakers that were not present in the training set.

% EXPAND: The field of multi-task learning has shown that training learned features on multiple separate but related tasks, can improve performance [cite]. 

% Comparison to meta learning literature: training a good feature extractor, classification head is just acting upon a good universal feature extractor. MAML Loop comparison.

% Discuss other embedding learning techniques - triplet loss, angular penalty losses, GE2E loss, centroid
Although the approach of achieving speaker-discriminative embeddings through training via classification is common \cite{Xie2019, Tang2019}, there exist other means in which to achieve this. For example, there are several approaches that are variants on triplet loss \cite{Hoffer2015,Heigold2016,Wan2017}, which explicitly optimizes embeddings to move closer to same-class examples whilst moving further away from out-of-class examples. Another approach is the family of angular penalty loss functions \cite{Wanga, Deng2018, Liu, Zhang2019}, which are similar to the standard softmax loss, but enforce a stricter condition on the decision boundary between classes by adding angular penalty terms for the correct class, thus encouraging larger intra-class distances and more compact inter-class distances.

We propose and make available code\footnote{\url{https://github.com/cvqluu/dropclass_speaker}} for two methods aimed at achieving speaker-discriminative embeddings, both focused around the notion of dropping classes during training. The first technique, referred to as DropClass, continually changes the training objective for deep speaker embedding systems by periodically dropping a random subset of classes during training, such that the network is continually trained on many different classification tasks. This is conceptually similar to applying Dropout \cite{Srivastava2014} on the output layer of a classification network while also disallowing training examples from the dropped classes. We argue that speaker recognition tasks have strong parallels with few-shot learning tasks and thus may benefit from a meta-learning style approach, which is what DropClass provides.

% We show that this technique can lead to a 7.9\% relative improvement in equal error rate (EER) over a strong baseline on the VoxCeleb speaker verification task \cite{Nagraniy2017}, and also results in gains on the Speakers In The Wild (SITW) core-core task \cite{Mclaren2016}.

%  We also show that a crucial component for this technique's success is the use of an angular penalty loss function [cite]
The second method that is proposed in this work, referred to as DropAdapt, can be applied to adapt a fully trained model to a set of enrolment speakers in an unsupervised manner. This is achieved by dropping the training classes which are are predicted by the model to be unlikely in the full set of enrolment utterances, rather than dropping randomly selected subsets of classes. We also show that the predicted distribution of speakers in the training set and test set can be heavily mismatched, which we argue negatively impacts performance. Our experiments show that DropAdapt can mitigate this distribution mismatch and that this correlates with improved speaker verification performance.

% When applied to VoxCeleb, this results in a large 13.2\% improvement in EER, with more modest improvements for SITW. We also demonstrate that the distribution of predicted class probabilities is far from uniform on the test set compared to the training set, and that the Kullback–Leibler (KL) divergence from the test set average predicted probability distribution to the uniform distribution has some correlation with the verification performance.
% Easily implemented to existing training

% to produce a robust feature extractor for speaker identity
\section{Dropping Training Classes}

\begin{figure*}[tb]
  \includegraphics[width=\textwidth]{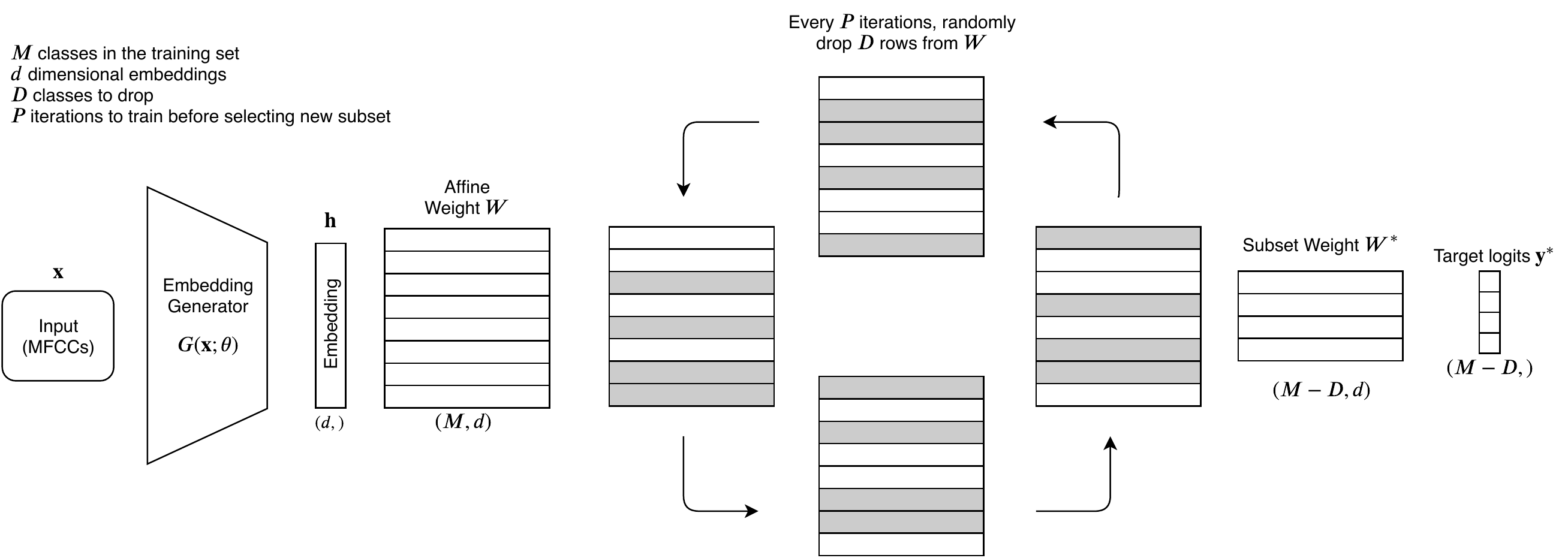}
  \caption{\label{fig:dc_diag}System diagram displaying the process of how classes are dropped throughout training in the proposed DropClass method.}
\end{figure*}

% \begin{itemize}
%     \item Motivation: diversify learning objective, encourage discriminability across multiple classification objectives
%     \item Periodically select a random subset of classes to include during training. This means only utterances belonging to the subset of allowed training classes are provided. This selected subset also determines which rows of the final affine transformation are included and thus are included in the classification softmax.
%     \item Throughout training, after $P$ iterations, $D$ classes are dropped, leaving $N - D$ classes for that period.
% \end{itemize}

% TODO: Expand this with set notation, more equations etc.

This work proposes two functionally similar techniques which revolve around the process of dropping classes during training. Despite this functional similarity, they have distinct applications, justifications, and links to literature which will be detailed in the following sections.

\subsection{DropClass}\label{sec:dropclass}

A typical architecture of a speaker embedding network, such as the successful x-vector architecture \cite{Snyder2018}, has the following structure. This network is trained as a whole, but can be split into two components, $G$ and $C$, which will be detailed below.

From an input $\bm{x}$ of acoustic features such as MFCCs, the first part of the network $G$ parameterized by $\theta$ can produce a $d$-dimensional embedding $\bm{h}$.

\begin{equation}
    \bm{h} = G(\bm{x};\theta)
\end{equation}

\noindent This is typically trained on a classification task, meaning the classification part of the overall network $C$, parameterized by $\phi$ acts on $\bm{h}$ to produce a prediction $\bm{y}$ for what class the input $\bm{x}$ belongs to.

\begin{equation}
    \label{eq:clf_eq}
    \bm{y} = C(\bm{h}; \phi)
\end{equation}

\noindent Both $C$ and $G$ are trained as a whole, usually via the standard cross entropy loss against a target one-hot vector $\hat{\bm{y}}$ which indicates which class out the set of $M$ training classes, $\mathcal{N}:\{i,\ldots,M\}$, $\bm{x}$ belongs to.

For simplicity, the classification network $C$ may be as rudimentary as an affine transform that projects the input embedding $\bm{h}$ into the correct number of dimensions, $M$. In this simplified case, the entirety of $\phi$ is a weight matrix $W$ with the dimensions $(M, d)$. Without a bias term, this changes Equation \ref{eq:clf_eq} to the following:

% Let us consider the classification layer of a deep embedding network (without a bias term), which applies an affine transformation to a generated embedding $\bm{x}$ of dimension $d$

\begin{equation}
    \label{eq:affine}
    \bm{y} = \bm{h}W^T
\end{equation}

\noindent where $\bm{y}$ contains the logits of the class prediction.

The proposed technique, referred to as DropClass, is detailed in Algorithm \ref{alg:dc}. When training with DropClass,   every $P$ iterations, a random subset of $\mathcal{N}$ is chosen: $\mathcal{R} \subset \mathcal{N}$ with size $M-D$ where $D$ is a variable that determines how many classes should be dropped. $P$ and $D$ are configurable hyperparameters.  The set $\mathcal{R}$ defines the permitted classes in the next $P$ iterations. The rows of the weight matrix $W$ which correspond to the subset of classes in $\mathcal{R}$ are selected to make a new matrix, $W^*$, which has the dimensions $(M-D, d)$, and  the output of the resulting modification of Equation \ref{eq:affine}, $\bm{y*}$ has dimension $(M-D)$:
\begin{equation}\label{eq:dropaffine}
    \bm{y^*} = \bm{h}{W^*}^T
\end{equation}
After $P$ iterations, the process is repeated and a new proper subset is randomly selected, with the process  continually repeated until training is completed (Figure \ref{fig:dc_diag}).

\begin{algorithm}[t]
\SetAlgoLined
\KwResult{Model trained with DropClass}
 \textbf{Given:} Feature extractor $G(\bm{x};\theta)$, Classification affine matrix $W$\\
 Set of all training classes $\mathcal{N}:\{i,\ldots,M\}$\\
 $D$ classes to drop per $P$ iterations\\
 Training dataset $\mathcal{D}_{\text{train}}$\\
 \While{not done}{
    Randomly sample proper subset of size $(M-D)$ from $\mathcal{N}$,
    \ $\mathcal{R}:\{j,\ldots,M-D\}\subset	\mathcal{N}$ \\
    % $\mathcal{R} \leftarrow \mathcal{N} - \mathcal{A}$ \\
    $W^{*} \leftarrow W[\text{Class rows in }\mathcal{R}]$\\
    $\mathcal{D}_{\text{temp}} \leftarrow \mathcal{D}_{\text{train}}[\text{Examples from classes in }\mathcal{R}]$  \\
  \For{$P$ iterations}{
    Train $G(\bm{x};\theta)$ and $W^*$ using $\mathcal{D}_{\text{temp}}$
    }
 }
 \caption{\label{alg:dc} DropClass approach to training a deep feature extractor.}
\end{algorithm}

This proposed method can be compared with a number of existing techniques in literature, in particular Dropout \cite{Srivastava2014}.  DropClass essentially drops units in the output classification layer and synchronizes this with the data provided to the model, ensuring that no dropped classes are provided while the corresponding classification units are dropped. 

% Performing dropout in this manner is not commonplace for classification tasks, perhaps due to 

% (Do I need to explain what Dropout is?) 
The effectiveness of Dropout has been justified by the technique performing a continuous sampling of an exponential number of thinned networks throughout training and then taking an average of these at test time \cite{Baldi2014,Warde-Farley2014}. As a result of this model averaging, Dropout has been shown to reduce overfitting and generally improve performance \cite{Srivastava2014}, and has seen widespread adoption in many different applications of neural networks \cite{Dahl2013, Variani2014, Wang2017a}. Similar in its justification, DropClass is continuously sampling from a large number of different classification tasks on which the embedding generator $G$ must perform well, in theory making it agnostic to any one specific task.

This technique also has some similarity to some techniques in the field of meta-learning for few-shot learning, specifically Model-Agnostic Meta Learning (MAML) \cite{Finn2017} and the related technique Almost No Inner Loop (ANIL) \cite{Raghu}. MAML is a method for tackling few-shot learning problems by utilizing two nested optimization loops. The outer loop finds an initialization for a network  which can adapt to new tasks quickly, whilst the inner loop uses the initialization from the outer loop and learns from a small number of examples from each desired task (referred to as the `support set'), performing a few gradient updates. 

Raghu et al \cite{Raghu} found the strength of MAML lay in the quality of the initialization found by the outer loop, with each task specific adaptation in the inner loop mostly reusing features already learned in the outer loop step. They proposed ANIL, which reduces the inner task-specific optimization loop to only optimize the classification layer, or `head', of a MAML-trained network. Similar to DropClass, ANIL makes a distinction between the part of the overall classification network which generates discriminative features (referred to as the `body'), and the classification head, which is more task specific. Raghu et al also proposed the No Inner Loop (NIL) method, which uses the cosine similarity between the generated features of an unseen example to the generated features of a small number of known examples to weight the classification prediction. This use of cosine similarity to compare embeddings is extremely commonplace in speaker recognition  \cite{Hansen2015} and in practice, the inference step of the NIL technique is identical to a $1$ to $N$ speaker identification set up, if one considers the utterances from the $N$ enrolment speakers to be the small number of labeled examples, the `support set'.

This similarity of the problems of the few-shot learning and speaker recognition tasks has influenced the proposal of DropClass, both of which aim to produce a `body' that generates features applicable to a distribution of tasks (sub-set classification) rather than to a single task. However, DropClass does not perform the outer and inner loops found in MAML/ANIL which explicitly optimizes the network to be robust to additional gradient steps per sub-task. Instead, DropClass encourages performance on all tasks by continually randomizing the training objective, implicitly encouraging the generated features to perform well across subtasks. Despite this, exploring ANIL and MAML for speaker representation learning would be a natural extension to this work. This extension would be particularly interesting considering the experiments on the NIL method (cosine similarity scoring) from Raghu et al \cite{Raghu}, specifically Table 5. They found that MAML and ANIL trained models significantly outperformed `multiclass training' models, where all possible classes were trained simultaneously. Considering the `multiclass training' paradigm is the most common approach to training deep speaker embedding extractors, there could well be gains to be found in adopting a meta-learning approach to training speaker embedding extractors.

\subsection{DropAdapt}\label{sec:dropadapt}

Deep speaker representations are optimized to distinguish between the training set speakers, which is hoped to generalise to any given set of new unseen speakers. Generally however, the distribution of speakers in a desired held out evaluation set does not exactly match the distribution seen during training; that is,  the expected distribution along the manifold of known speakers is often not replicated in the evaluation set. A clearer explanation for this can be seen if we examine what classes are predicted by the whole network when we give as input the utterances in the test set. 

% These predictions, whilst not particularly instructive in terms of discriminating between speakers, do give insight as to 
Starting from a trained model, for a given dataset $\mathcal{D}$ of $N$ examples, the average probability assigned to each class can be calculated as follows,

\begin{equation}\label{eq:p_avg}
    \bm{p}_{\text{average}} = \frac{1}{N} \sum^{N}_{i=1} \softmax ( \bm{h}_i W^T ) 
\end{equation}

\noindent where $\bm{h}_i$ is the embedding extracted from the $i^{\text{th}}$ utterance, and $W$ is the final affine weight matrix. The resulting $M$-dimensional vector $\bm{p}_{\text{average}}$ is a representation of the mean probability that the model predicts for the presence of each speaker across the $N$ utterances. Provided a uniform distribution of speakers was used to train the model, it would be expected that the model predict a near uniform $\bm{p}_{\text{average}}$ for an input of training examples with uniform class distribution. This however may not be the case if the model is provided a selection of evaluation utterances with uniform speaker distribution.

\begin{figure}[t]
\includegraphics[width=\columnwidth]{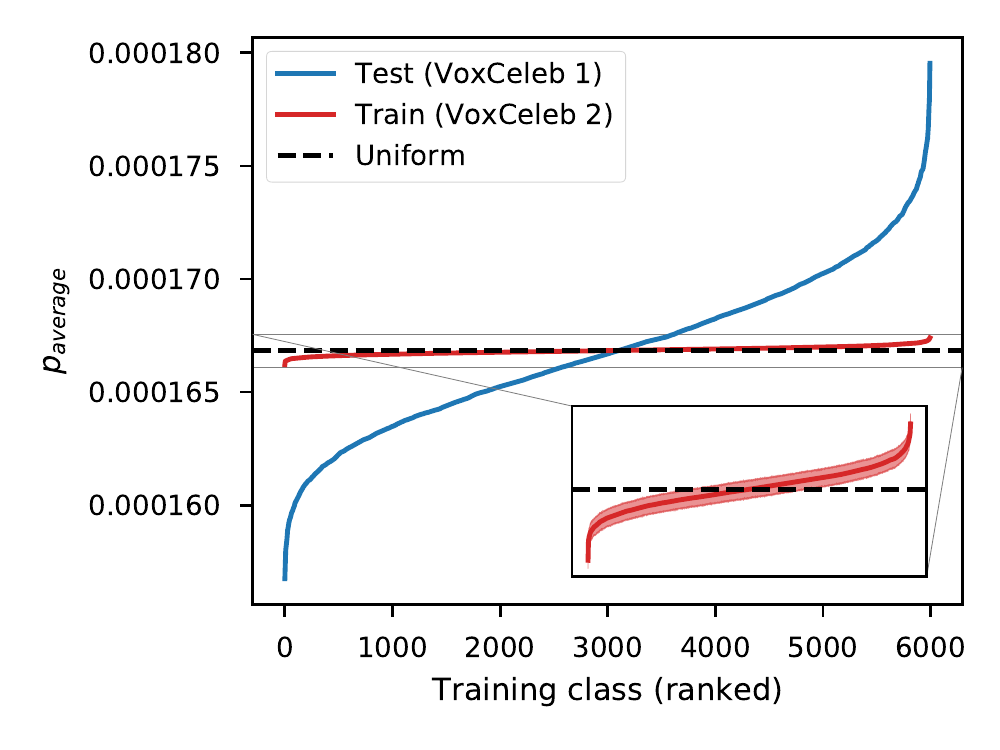}
\vspace{-5mm}
\caption{\label{fig:vcprobskew}{Comparison of the 5994 ranked training class probability predictions from a training set and test set, both provided with 40 unique speakers with 42 examples each. The training set probability has uncertainty bounds for 300 bootstrap sampled variations of speakers and examples.}}
\end{figure}

This effect can be seen in Figure \ref{fig:vcprobskew}, where a trained model predicts close to the uniform distribution of classes when provided a uniform class distribution of training examples, but predicts a much more skewed $\bm{p}_{\text{average}}$ on the test set, with some training classes predicted to be much more likely than others. This is perhaps not a surprising result, as in the hypothetical situation of a test set with entirely one gender, a skewed $\bm{p}_{\text{average}}$ is expected. However, this skew is often not as clearly explainable as the hypothetical one-gender test set, and may have multiple contributing factors. For context, the VoxCeleb 1 test set has a `good balance of male and female speakers' \cite{Nagraniy2017}, and the speakers in it were chosen because their names began with the letter `E'.

This observation can be interpreted in a number of ways. For example, it is well known that class imbalance is a significant impedance to performance in classification tasks \cite{Buda2017}, especially in cases in which training and inference have significantly different distributions. It is a natural extension to this that the performance of an embedding extracted from a classification network would degrade in performance in the same manner, which has been seen in the work of Huang et al and Khan et al \cite{Huang,Khan2018}. In these works, they found cost sensitive training and oversampling methods to increase the performance of learned representations.

% EXPAND (more huang papers Storkey paper) \cite{Storkey2013WhenOverview}.
The second closely related interpretation and hypothesis is that the `low probability' classes predicted by the model are in some way less important to the performance of the embeddings on the test set. These `low probability' classes are suggested by the model's predictions to be less likely to be present in the test set. This might imply that distinguishing between these specific classes is not crucial to the end task.

Following from these interpretations, this technique, referred to as DropAdapt, works via dropping these low probability classes permanently to fine-tune a fully trained model, adapting the model to a test set and hopefully increasing performance. This is described in Algorithm \ref{alg:dca}. This method in theory should be applied only to a fully or near fully trained model, as an accurate estimation of the training class occupation must be obtained first.

To ensure an accurate probability estimation of the test set throughout the fine-tuning, this ranking (and dropping) of the least probable classes can be performed periodically, meaning this technique is functionally similar to the DropClass method above, except that classes are removed permanently, and the dropping of classes is determined by the probability criterion $\bm{p}_{\text{average}}$ instead of randomly.

A slight variation of this method explored in this work is referred to DropAdapt-Combine, in which instead of permanently removing these classes, all the low probability classes are combined into a single new class such that the examples belonging to the removed classes are not completely discarded.

\begin{algorithm}[t]
\SetAlgoLined
\KwResult{Model tuned with DropAdapt}
 \textbf{Given:} Trained feature extractor $G(\bm{x};\theta)$, trained $W$\\
 Training set $\mathcal{D}_{\text{train}}$\\
 Unlabeled Test/enrolment utterances $\mathcal{D}_{\text{enrol}}$\\
 \While{not done}{
    Calculate all $\bm{p}_{\text{average}}$ from $\mathcal{D}_{\text{enrol}}$ (see Equation \ref{eq:p_avg})\\
    Rank classes by $\bm{p}_{\text{average}}$ \\
    Select set of higher probability classes from $\mathcal{N}$, dropping lowest probability $D$ classes $\rightarrow \mathcal{R}$\\
    \eIf{Combine}{
       Assign all examples not in $\mathcal{R}$ same class label in $\mathcal{D}_{\text{train}}$\\
       $W^* \leftarrow W[\text{Class rows in }\mathcal{R}]$ \\
   }{
        $\mathcal{D}_{\text{train}} \leftarrow \mathcal{D}_{\text{train}}[\text{Examples from classes in }\mathcal{R}]$\\
        $W^* \leftarrow W[\text{Class rows in }\mathcal{R}]$ \\
  }
  $W \leftarrow W^*$\\
  $\mathcal{N} \leftarrow \mathcal{R}$\\
    % $\mathcal{R}:\{j,\ldots,M-D\}\subset \mathcal{N}$ \\
    % $W^{*} \leftarrow W[\text{Class rows in }\mathcal{R}]$\\
  \For{$P$ iterations}{
    Train $G(\bm{x};\theta)$ and $W$ using $\mathcal{D}_{\text{train}}$
    }
 }
 \caption{\label{alg:dca}DropAdapt method for adapting a deep feature extractor to a chosen dataset}
\end{algorithm}

This method can be compared to techniques in the fields of active learning and learning from small amounts of data, such as the Facility-Location and Disparity-Min models \cite{Kaushal2019}, which put heavy emphasis on selecting the right subset of examples in order to learn efficiently. These methods are typically used to capture the whole distribution of the desired dataset in as few examples as possible, encouraging a diverse and representative subset of examples. However, it is implied by Figure \ref{fig:vcprobskew} that in this speaker embedding task, even if the whole training dataset were used, this may not be representative of the distribution found at test time. DropAdapt can be seen as a means of correcting this mismatch through subset selection for fine-tuning. 

Buda et al \cite{Buda2017} and Huang et al. \cite{Huang} found oversampling minority classes to be an effective strategy in improving performance for neural networks on imbalanced datasets. Viewing this problem as a dataset imbalance problem, DropAdapt could also be interpreted as a corrective oversampling strategy, training additionally on those classes which are retained to better match the target distribution.

% training additionally on classes which are, according to the model, underrepresented in the test set compared to the training set.

This train-test distribution mismatch is also closely linked to the field of domain adaptation and the domain-shift problem \cite{Patel2015}. However, DropAdapt is primarily proposed as a means of adapting to a class distribution mismatch, as it is likely that $\bm{p}_{\text{average}}$ is less informative the greater the domain mismatch. Combining domain adaptation techniques with DropAdapt could be an interesting extension to this work.

% label smoothing (uncertainty for a single classification problem)?
% EXPAND: Active learning literature, dataset shift techniques.

% \begin{itemize}
%     \item Motivation: enrolment speakers in test scenario are more important to distinguish (not necessarily between them)
%     \item Starting from a fully trained model
%     \item Aggregate output probabilities of training classes but on the enrolment (test set) utterances
%     \item Select desired number of low probability classes (consider these to be not as relevant for the desired task)
%     \item Drop these classes and train again. Can repeat this step, iteratively removing classes.
%     \item Either can remove low prob speakers completely, or combine low probability classes into a single class - motivation: unknown speakers)
% \end{itemize}

% \begin{itemize}
%     \item Only VoxCeleb 2 used as training dataset (5994 speakers)
%     \item 30-dim MFCCs
%     \item standard Kaldi data aug routine (noise, music, babble, reverb)
%     \item X-vector arch (angular penalty applied directly to embedding layer)
%     \item additive margin penalty
%     \item batch size 500
%     \item 120000 iterations
%     \item SGD, lr 0.2, mom 0.5
%     \item Halving learning rate at 60000, 80000, 90000, 110000 steps
%     \item Ensure each batch has same number of unique speakers as examples
%     \item Embeddings L2 normalized and compared with Cosine distance
% \end{itemize}

\section{Experiments}

The following section details the experimental setup and the experiments performed utilizing the proposed methods.

\subsection{Experimental Setup}

The primary task that these experiments attempted to improve performance on was that of speaker verification, specifically that on VoxCeleb 1 \cite{Nagraniy2017} and Speakers In The Wild (SITW) core-core task \cite{Mclaren2016}. Although there exist several metrics to evaluate verification performance, which are typically chosen depending on the desired behaviour of a system, the primary metric explored here was the equal error rate (EER), as that is the primary metric for evaluation on VoxCeleb 1.

The training data used for all experiments was the VoxCeleb 2 development set \cite{Chung2018}, which features 5994 unique speakers. This was augmented in the standard Kaldi\footnote{https://kaldi-asr.org/} fashion with noise, music, babble and reverberation. The original x-vector architecture was used with very little modification, using Leaky ReLU instead of ReLU, with $30$-dimensional MFCC features as inputs, and 512-dimensional embeddings. The main difference between this implementation and that of Snyder et al \cite{Snyder2018} was the use of the CosFace \cite{Wang2018} angular penalty loss function instead of a traditional cross entropy loss. This classification transform also was applied directly to the embedding layer, unlike the original, which has an additional hidden layer between the embedding layer and the classification layer. This means that the simplified notation for the classifier $C$ following from equation \ref{eq:affine} is an accurate representation of our model. All pairs of embeddings were $L2$ normalized and scored using cosine distance.

A batch size of 500 was used, with each example having 350 frames. Each batch had the same number of unique speakers as examples. Models were trained for 120,000 iterations, using SGD with a learning rate of 0.2 and momentum 0.5. The learning rate was halved at 60,000, 80,000, 90,000, and 110,000 steps. For DropAdapt fine-tuning, the learning rate was chosen to be the same as it was at the end of training the original model, and all the enrolment utterances were used to calculate $\bm{p}_{\text{average}}$.

\subsection{DropClass Experiments}

\begin{figure}[t]
\includegraphics[width=\columnwidth]{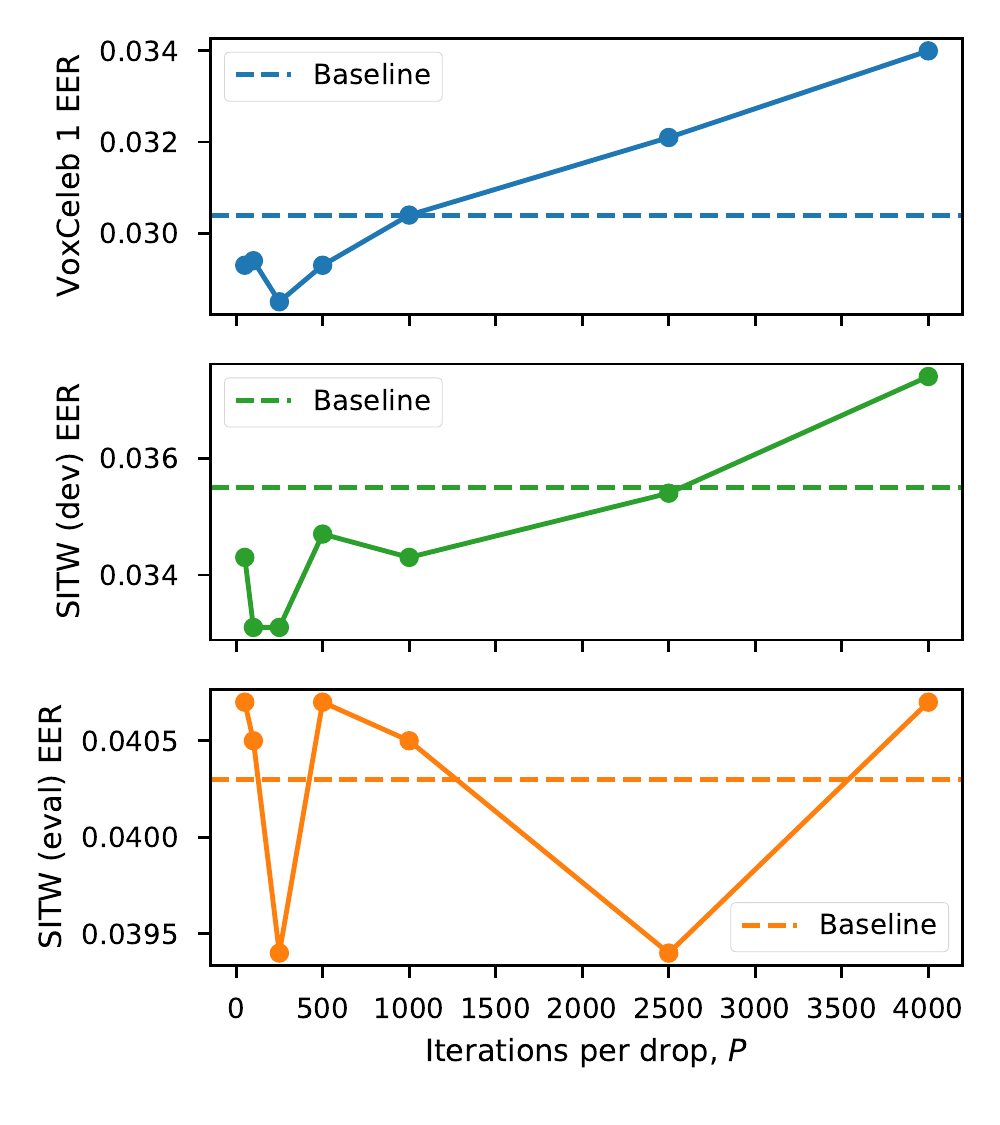}
\vspace{-5mm}
\caption{\label{fig:dc_its}{Comparison on the effect on EER of varying the number of iterations to run before re-selecting the class subset, fixed at dropping $5000/5994$ training classes each period.}}
\end{figure}

\begin{figure}[t]
\includegraphics[width=\columnwidth]{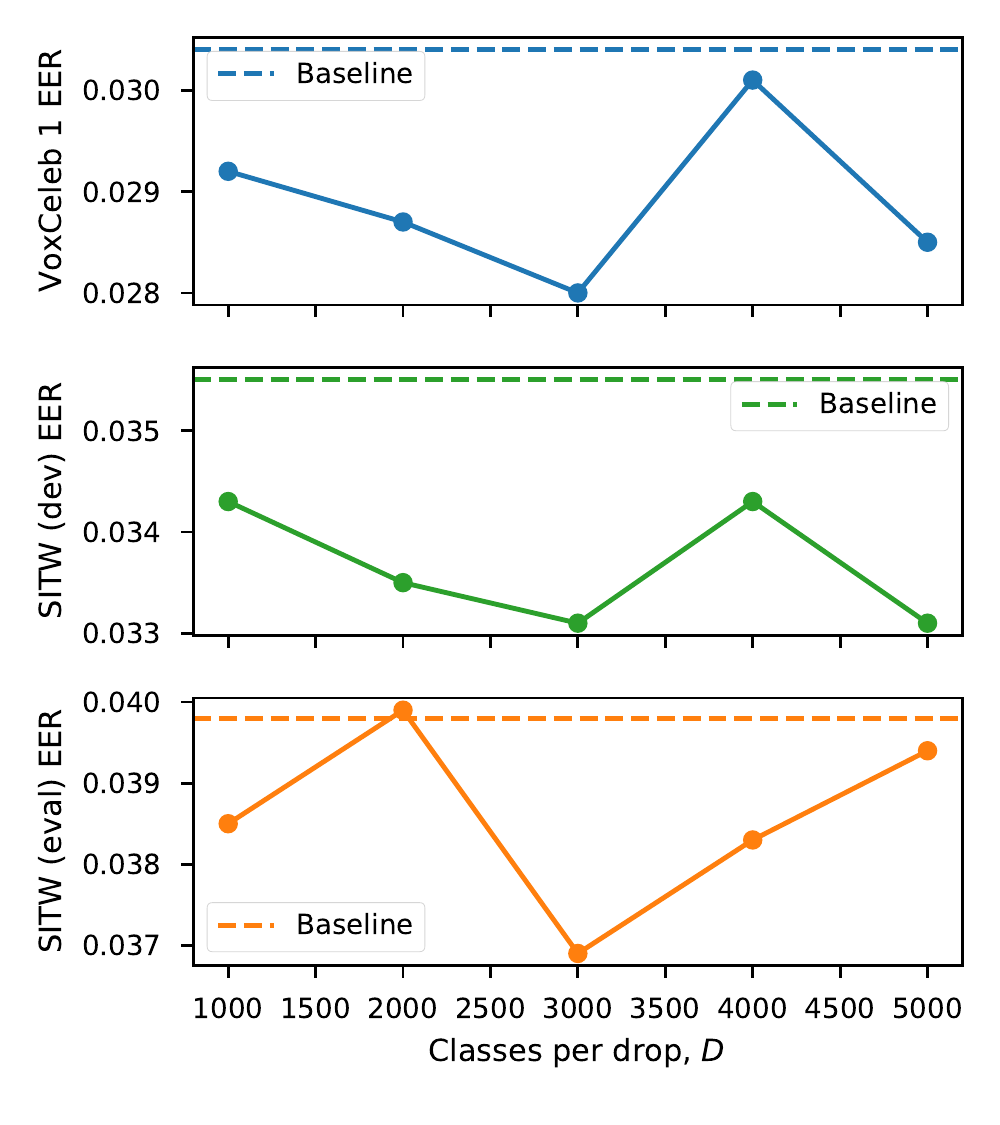}
\vspace{-5mm}
\caption{\label{fig:dc_cls}{Comparison on the effect on EER of varying the number of classes to cycle out every $250$ iterations.}}
\end{figure}

% MAKE THIS WHOLE SECTION LESS DISJOINTED:
Our initial experiments investigated favourable settings of
$P$ and $D$ for DropClass, and the results are shown in Figure \ref{fig:dc_its}, where the number of classes to drop was fixed at $D=5000$ and the number of iterations $P$ was varied between 50 and 4,000, the latter being slightly over 1 epoch's worth of data with the chosen batch size. It can be seen that improvements over the baseline are to be found more reliably at lower values of $P$, with consistent performance improvements when $P< 1000$ for both VoxCeleb and SITW (dev). This is perhaps unsurprising, as a motivating factor for this technique was to train a network on many different permutations for robustness on a variety of tasks, and thus with a training budget for each model of 120,000 iterations, this is not a large number of permutations throughout training. As the value for $P$ increases, this may also increase the risk of incurring the phenomenon of catastrophic forgetting \cite{French1999,Goodfellow2013}, an issue in which networks trained on a new task begin to degrade on the task that they previously were trained on.

From the previous experiment, choosing the best performing value of $P=250$ across each dataset, the number of classes to drop $D$ was then varied from 1000 to 5000, shown in Figure \ref{fig:dc_cls}. From this, we can see that for nearly all configurations of $D$, performance was improved on all datasets using DropClass over the baseline. Dropping approximately half the classes at $D$=3000 appeared to produce the best performance, although a more thorough exploration with different training data is likely required to ascertain if any heuristic exists for the selection of this value. However, from the previous experiments, it can be seen that for a suitably low value of $P$, DropClass can convey improvements over the baseline.

% The best configuration for all datasets was the model trained with $P$=250, $C$=3000. 
It is however important to note that a crucial component of this method is the use of the CosFace \cite{Wang2018} angular penalty loss, with Table \ref{tab:loss_func} showing a comparison of the effect that changing the loss function had on the improvement that DropClass produced on VoxCeleb. A more in-depth analysis on how each loss function changes with the permutations of each subset of classes is required.

\subsection{DropAdapt Experiments}

Table \ref{tab:dca_eer} displays the relative improvement in EER from utilizing the DropAdapt and DropAdapt-Combine method, using either the enrolment speakers from VoxCeleb 1 or SITW (dev) to choose which speakers to drop. The starting point was a standard classification trained baseline. Models were trained on a budget of 30,000 iterations, and one configuration for $D$ and $P$ was tested. Also compared were the following control experiments: The baseline but trained additionally for the same number of iterations as DropAdapt, Drop-Random, which drops random classes permanently, ignoring the $\bm{p}_{\text{average}}$ score, and Drop Only Data, which removes the low probability classes from the training data, but does not remove the relevant rows in the final weight matrix, bypassing the use of $W^*$ in Equation \ref{eq:dropaffine}.

Compared to the baseline and control experiments, both DropAdapt and DropAdapt-Combine show strong performance gains on VoxCeleb. The $2.63\%$ EER on VoxCeleb is particularly impressive when compared to other works which use similar or larger network architectures and more training data and achieve $>3\%$ EER \cite{Okabe2018,Xie2019}. The improvements over the baseline on SITW however are more modest, with DropClass trained models and `Drop Only Data' outperforming the DropAdapt models. 

An interesting observation is the fact that dropping only the data improved performance on VoxCeleb, but not as much as the DropAdapt methods. As discussed in section \ref{sec:dropadapt}, DropAdapt can be viewed as a form of corrective oversampling of targeted classes, with oversampling techniques having been shown to improve performance in imbalanced data scenarios \cite{Huang, Khan2018}. From this, we can see that for the within-domain data, some of the benefit of DropClass is gained from only fine-tuning via oversampling, but this benefit is increased further by also dropping the classes from the output layer. Conversely, for the out-of-domain SITW dataset, dropping only the classes from the data performed the best. We hypothesize that the reduced effectiveness of DropAdapt in this case may be due to the technique having to adapt to not only a new speaker distribution, but also a new domain. Further exploration combining DropAdapt with traditional domain adaptation techniques is left for future work.

% We hypothesize that the feature transformation caused by DropAdapt is larger than only 

% It may be for example that the class distribution predicted by the model on out of domain data is not accurate, with the domain shift affecting this estimation. 
In addition, more experimentation on the configurations of $P$ and $D$ could be explored, as it may be possible for example that the iterative dropping of classes is not necessary, and that the initial probability estimation is suitable. Furthermore, the most obvious extension left for future work is to use both DropClass and DropAdapt in conjunction, as both have been shown to provide performance increases in parallel.

\begin{figure}[t!]
\includegraphics[width=\columnwidth]{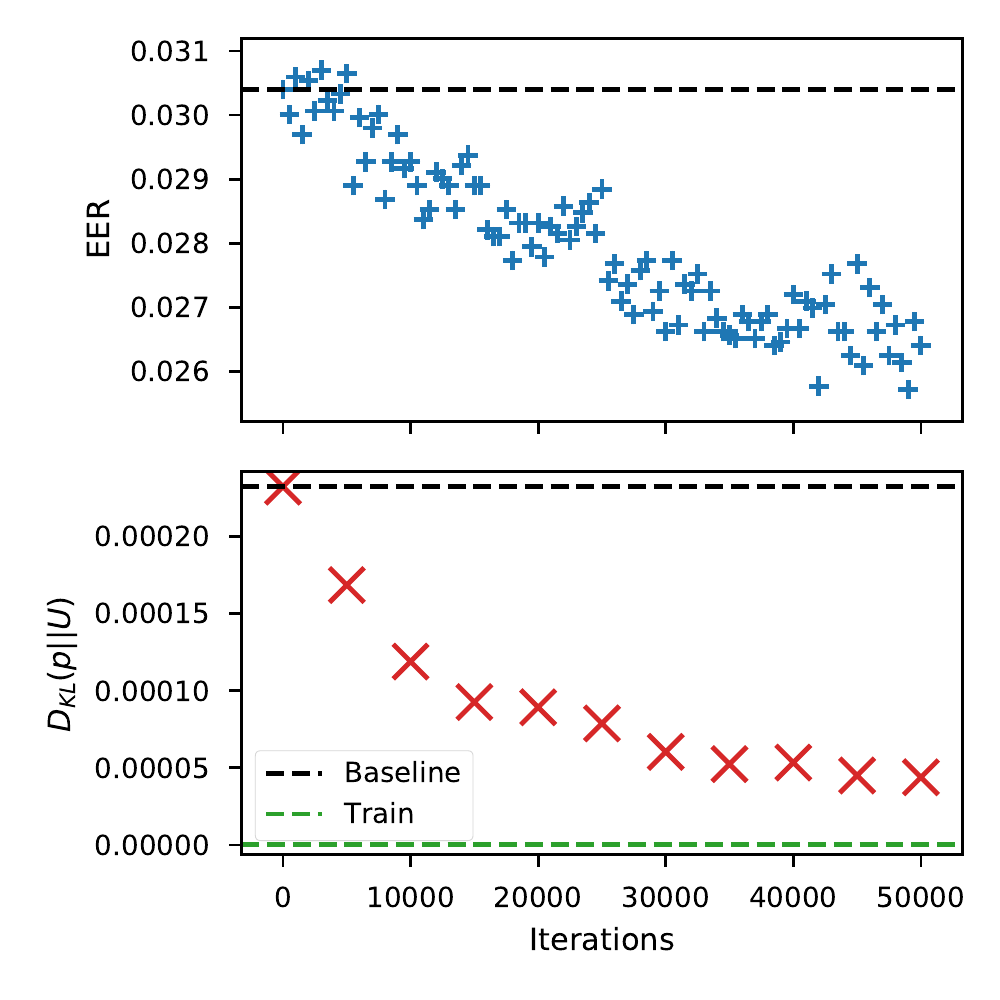}
\vspace{-10mm}
\caption{\label{fig:dc_entropy}Plot of the evolving EER on VoxCeleb as classes are dropped with DropAdapt-Combine ($P$=5000, $D$=500) along with the KL divergence from $\bm{p}_{\text{average}}$ on the test set to the uniform distribution.}
\end{figure}

Following up on the hypothesis presented in section \ref{sec:dropadapt} that the imbalanced distribution of $\bm{p}_{\text{average}}$ on the test set may be an indicator of train-test mismatch and thus incurring performance loss, Figure \ref{fig:dc_entropy} shows the EER and the KL divergence ($D_{KL}(p||U)$) from the VoxCeleb test set $\bm{p}_{\text{average}}$ to the uniform distribution as the DropAdapt-Combine model is trained. As we can see from the figure, while the EER decreases, the distribution of $\bm{p}_{\text{average}}$ also gets closer to the uniform distribution. Whilst there appears to be a correlation, this is likely not a strongly linked pair of observations, in that we can easily break this relationship by training only the final affine matrix $W$ and freezing the embedding extractor to provide more favourable class weightings for $\bm{p}_{\text{average}}$. However, in the case of DropAdapt, the decreasing $D_{KL}(p||U)$ may indicate that a favourable change in the extracted representations is occurring. This could be useful as a stopping criterion for cases in which adaptation data has no labels at all.

% \begin{itemize}
%     \item Figure \ref{fig:dc_its} shows the first exploration of the proposed technique, dropping $D=5000$ classes and varying the reset period $P$. 
%     \item Lower iterations (below $1000$) seem to perform well, outperforming the baseline on VC1, $250$ iterations is the only configuration which outperforms the baseline in all datasets.
%     \item 
% \end{itemize}

\begin{table}[t!]
  \centering
  \begin{tabular}{| m{3cm} || c c |}
  \hline
  & \multicolumn{2}{c|}{EER (VoxCeleb)} \\
  \cline{2-3}
    & Baseline & DropClass \\ 
  \hline
   Softmax & 5.89\% & 6.25\% \\
   CosFace \cite{Wang2018} & \textbf{3.04\%} & \textbf{2.80\%}  \\
   SphereFace \cite{Liu} & 3.92\% & 4.76\%  \\
   ArcFace \cite{Deng2018} & 3.19\% & 3.08\%  \\
   AdaCos \cite{Zhang2019} & 3.24\% & 3.81\%  \\
   \hline
  \end{tabular}
  \caption{\label{tab:loss_func}EER values on VoxCeleb 1 for using DropClass ($P$=250, $D$=3000) or not with different angular penalty loss functions, all with the paper recommended settings of hyperparameters.}
\end{table}

\begin{table}[t!]
  \centering
  \begin{tabular}{| m{4.65cm} || c c |}
  \hline
    & EER & Rel Impr \\ 
  \hline
   Baseline (VoxCeleb) & 3.04\% & - \\
   Baseline (More iterations) & 3.06\% & -0.7\% \\
   Drop Random ($D$=500, $P$=5000) & 3.08\% & -1.3\% \\
   Drop Only Data ($D$=500, $P$=5000) & 2.86\% & 5.9\% \\
   DropAdapt ($D$=500, $P$=5000) & 2.68\% & 11.8\% \\
   DropAdapt-C ($D$=500, $P$=5000) & \textbf{2.64\%} & \textbf{13.2\%} \\
   \hline
   Baseline (SITW) & 3.55\% & - \\
   Baseline (More iterations) & 3.61\% & -1.7\% \\
   Drop-Random ($D$=500, $P$=5000) & 3.73\% & -5.1\% \\
   Drop Only Data ($D$=500, $P$=5000) & \textbf{3.31\%} & \textbf{6.7\%} \\
   DropAdapt ($D$=500, $P$=5000) & 3.47\% & 2.3\% \\
   DropAdapt-C ($D$=500, $P$=5000) & 3.39\% & 4.5\% \\
   \hline
  \end{tabular}
  \caption{\label{tab:dca_eer} Relative improvment in EER from using DropAdapt and DropAdapt-Combine (DropAdapt-C) on the VoxCeleb 1 and SITW datasets on a budget of $30,000$ iterations}
\end{table}

\section{Conclusion}

In this work we presented the DropClass and DropAdapt methods for training and fine-tuning deep speaker embeddings. Both methods are based around the notion of dropping classes from the final classification output layer while also withholding examples belonging to those same classes. Drawing inspiration from Dropout and meta-learning, DropClass is a method that drops classes randomly and periodically throughout training such that a model is trained on a large number of different classification objectives for subsets of the training classes as opposed to classifying on the full set of classes. We show that in conjunction with the CosFace \cite{Wang2018} loss function, DropClass can improve verification performance on the VoxCeleb and SITW core-core tasks. 

We present the mismatch in class imbalance between train and test as a potential reason for reduced performance in verification, and propose DropAdapt as a means of alleviating this. DropAdapt is a method which can adapt a trained model to a target dataset with unknown speakers in an unsupervised manner. This is achieved by calculating the average predicted probability of each training class with the adaptation data as input. From these predictions, the model is fine-tuned by dropping the low probability classes and training for more iterations, focusing on the classes which the model has predicted to be presented in the adaptation dataset. This is not unlike traditional oversampling techniques. Applying DropAdapt to VoxCeleb leads to a large improvement over the baseline, with DropAdapt also outperforming simply oversampling the same classes, suggesting it may be an effective strategy in adapting to a different class distribution than what was seen during training. We also show empirically that as the class distribution mismatch is corrected during DropAdapt, so too does the verification performance increase.

\bibliographystyle{IEEEbib}
\bibliography{newrefs.bib}

\begin{thebibliography}{10}

\bibitem{Snyder}
David Snyder, Daniel Garcia-Romero, Daniel Povey, and Sanjeev Khudanpur,
\newblock ``{Deep Neural Network Embeddings for Text-Independent Speaker
  Verification},''
\newblock in {\em Interspeech}, ISCA, 8 2017, pp. 999--1003, ISCA.

\bibitem{Snyder2018}
David Snyder, Daniel Garcia-Romero, Gregory Sell, Daniel Povey, and Sanjeev
  Khudanpur,
\newblock ``{X-vectors: robust DNN embeddings for speaker recognition},''
\newblock in {\em IEEE ICASSP}, 2018, pp. 5329--5333.

\bibitem{Dehak2011}
Najim Dehak, Patrick~J. Kenny, R{\'{e}}da Dehak, Pierre Dumouchel, and Pierre
  Ouellet,
\newblock ``{Front end factor analysis for speaker verification},''
\newblock {\em IEEE Trans. Audio. Speech. Lang. Processing}, vol. 19, no. 4,
  pp. 788--798, 2011.

\bibitem{Sell2018}
Gregory Sell, David Snyder, Alan McCree, Daniel Garcia-Romero, Jesús Villalba,
  Matthew Maciejewski, Vimal Manohar, Najim Dehak, Daniel Povey, Shinji
  Watanabe, and Sanjeev Khudanpur,
\newblock ``{Diarization is Hard: Some Experiences and Lessons Learned for the
  JHU Team in the Inaugural DIHARD Challenge},''
\newblock in {\em Interspeech}. 09 2018, pp. 2808--2812, ISCA.

\bibitem{Diez2019}
Mireia Diez, Lukas Burget, Shuai Wang, Johan Rohdin, and Honza Cernock{\'{y}},
\newblock ``{Bayesian HMM based x-vector clustering for Speaker Diarization},''
\newblock {\em Interspeech}, pp. 346--350, 2019.

\bibitem{Villalba2020}
Jes{\'{u}}s Villalba, Nanxin Chen, David Snyder, Daniel Garcia-Romero, Alan
  McCree, Gregory Sell, Jonas Borgstrom, Leibny~Paola Garc{\'{i}}a-Perera, Fred
  Richardson, R{\'{e}}da Dehak, Pedro~A. Torres-Carrasquillo, and Najim Dehak,
\newblock ``{State-of-the-art speaker recognition with neural network
  embeddings in NIST SRE18 and Speakers in the Wild evaluations},''
\newblock {\em Computer Speech and Language}, vol. 60, 2020.

\bibitem{Xie2019}
W.~{Xie}, A.~{Nagrani}, J.~S. {Chung}, and A.~{Zisserman},
\newblock ``Utterance-level aggregation for speaker recognition in the wild,''
\newblock in {\em IEEE ICASSP}, May 2019, pp. 5791--5795.

\bibitem{Tang2019}
Yun Tang, Guo-Hong Ding, Jing Huang, Xiaodong He, and Bowen Zhou,
\newblock ``Deep speaker embedding learning with multi-level pooling for
  text-independent speaker verification,''
\newblock {\em IEEE ICASSP}, pp. 6116--6120, 2019.

\bibitem{Hoffer2015}
Elad Hoffer and Nir Ailon,
\newblock ``Deep metric learning using triplet network,''
\newblock in {\em Similarity-Based Pattern Recognition}, Cham, 2015, pp.
  84--92, Springer International Publishing.

\bibitem{Heigold2016}
Georg Heigold, Ignacio Moreno, Samy Bengio, and Noam Shazeer,
\newblock ``{End-to-end text-dependent speaker verification},''
\newblock in {\em IEEE ICASSP}, 2016, vol. 2016-May, pp. 5115--5119.

\bibitem{Wan2017}
Li~Wan, Quan Wang, Alan Papir, and Ignacio~Lopez Moreno,
\newblock ``{Generalized End-to-End Loss for Speaker Verification},''
\newblock {\em arXiv}, 2017.

\bibitem{Wanga}
Feng Wang, Jian Cheng, Weiyang Liu, and Haijun Liu,
\newblock ``{Additive margin softmax for face verification},''
\newblock {\em IEEE Signal Processing Letters}, vol. 25, no. 7, pp. 926--930,
  2018.

\bibitem{Deng2018}
Jiankang Deng, Jia Guo, Niannan Xue, and Stefanos Zafeiriou,
\newblock ``Arcface: Additive angular margin loss for deep face recognition,''
\newblock in {\em IEEE CVPR}, June 2019, pp. 4690--4699.

\bibitem{Liu}
Weiyang Liu, Yandong Wen, Zhiding Yu, Ming Li, Bhiksha Raj, and Le~Song,
\newblock ``{SphereFace: Deep hypersphere embedding for face recognition},''
\newblock in {\em IEEE CVPR}, 2017, pp. 6738--6746.

\bibitem{Zhang2019}
Xiao Zhang, Rui Zhao, Yu~Qiao, Xiaogang Wang, and Hongsheng Li,
\newblock ``{AdaCos: Adaptively} scaling cosine logits for effectively learning
  deep face representations,''
\newblock in {\em IEEE CVPR}, 2019, pp. 10815--10824.

\bibitem{Srivastava2014}
Nitish Srivastava, Geoffrey Hinton, Alex Krizhevsky, Ilya Sutskever, and Ruslan
  Salakhutdinov,
\newblock ``{Dropout: A} simple way to prevent neural networks from
  overfitting,''
\newblock {\em J. Mach. Learn. Res.}, vol. 15, pp. 1929--1958, 2014.

\bibitem{Baldi2014}
Pierre Baldi and Peter Sadowski,
\newblock ``{The dropout learning algorithm},''
\newblock {\em Artif. Intell.}, vol. 210, no. 1, pp. 78--122, 2014.

\bibitem{Warde-Farley2014}
David Warde-Farley, Ian~J. Goodfellow, Aaron Courville, and Yoshua Bengio,
\newblock ``{An empirical analysis of dropout in piecewise linear networks},''
\newblock {\em ICLR}, pp. 1--10, 2014.

\bibitem{Dahl2013}
George~E. Dahl, Tara~N. Sainath, and Geoffrey~E. Hinton,
\newblock ``{Improving deep neural networks for LVCSR using rectified linear
  units and dropout},''
\newblock in {\em IEEE ICASSP}, 2013, pp. 8609--8613.

\bibitem{Variani2014}
Ehsan Variani, Xin Lei, Erik McDermott, Ignacio~Lopez Moreno, and Javier
  Gonzalez-Dominguez,
\newblock ``{Deep neural networks for small footprint text-dependent speaker
  verification},''
\newblock in {\em IEEE ICASSP}. 2014, pp. 4052--4056, Institute of Electrical
  and Electronics Engineers Inc.

\bibitem{Wang2017a}
Yuxuan Wang, R.~J. Skerry-Ryan, Daisy Stanton, Yonghui Wu, Ron~J. Weiss,
  Navdeep Jaitly, Zongheng Yang, Ying Xiao, Zhifeng Chen, Samy Bengio, and Quoc
  Le,
\newblock ``{Tacotron: Towards end-To-end speech synthesis},''
\newblock in {\em Interspeech}. 2017, vol. 2017-Augus, pp. 4006--4010,
  International Speech Communication Association.

\bibitem{Finn2017}
Chelsea Finn, Pieter Abbeel, and Sergey Levine,
\newblock ``{Model-agnostic meta-learning for fast adaptation of deep
  networks},''
\newblock in {\em ICML}, 2017, vol.~3, pp. 1856--1868.

\bibitem{Raghu}
Aniruddh Raghu, Maithra Raghu, Samy Bengio, and Oriol Vinyals,
\newblock ``{Rapid Learning or Feature Reuse? Towards Understanding the
  Effectiveness of MAML},'' 2019.

\bibitem{Hansen2015}
John~H.L. Hansen and Taufiq Hasan,
\newblock ``{Speaker recognition by machines and humans: A tutorial review},''
\newblock {\em IEEE Signal Process. Mag.}, vol. 32, no. 6, pp. 74--99, 2015.

\bibitem{Nagraniy2017}
Arsha Nagrani, Joon~Son Chung, and Andrew Zisserman,
\newblock ``{VoxCeleb: A large-scale speaker identification dataset},''
\newblock {\em Interspeech}, pp. 2616--2620, 2017.

\bibitem{Buda2017}
Mateusz Buda, Atsuto Maki, and Maciej~A. Mazurowski,
\newblock ``{A systematic study of the class imbalance problem in convolutional
  neural networks},''
\newblock {\em Neural Networks}, vol. 106, pp. 249--259, oct 2018.

\bibitem{Huang}
Chen Huang, Yining Li, Change~Loy Chen, and Xiaoou Tang,
\newblock ``Deep imbalanced learning for face recognition and attribute
  prediction,''
\newblock {\em IEEE Transactions on Pattern Analysis and Machine Intelligence},
  p. 1–1, 2019.

\bibitem{Khan2018}
Salman~H. Khan, Munawar Hayat, Mohammed Bennamoun, Ferdous~A. Sohel, and
  Roberto Togneri,
\newblock ``{Cost-sensitive learning of deep feature representations from
  imbalanced data},''
\newblock {\em IEEE Trans. Neural Networks Learn. Syst.}, vol. 29, no. 8, pp.
  3573--3587, 2018.

\bibitem{Kaushal2019}
V.~{Kaushal}, R.~{Iyer}, S.~{Kothawade}, R.~{Mahadev}, K.~{Doctor}, and
  G.~{Ramakrishnan},
\newblock ``Learning from less data: A unified data subset selection and active
  learning framework for computer vision,''
\newblock in {\em IEEE WACV}, Jan 2019, pp. 1289--1299.

\bibitem{Patel2015}
V.~M. {Patel}, R.~{Gopalan}, R.~{Li}, and R.~{Chellappa},
\newblock ``Visual domain adaptation: A survey of recent advances,''
\newblock {\em IEEE Signal Processing Magazine}, vol. 32, no. 3, pp. 53--69,
  May 2015.

\bibitem{Mclaren2016}
Mitchell McLaren, Luciana Ferrer, Diego Castan, and Aaron Lawson,
\newblock ``{The speakers in the wild (SITW) speaker recognition database},''
\newblock in {\em Interspeech}, 2016, vol. 08-12-Sept, pp. 818--822.

\bibitem{Chung2018}
Joon~Son Chung, Arsha Nagrani, and Andrew Zisserman,
\newblock ``{Voxceleb2: Deep speaker recognition},''
\newblock {\em Interspeech}, , no. ii, pp. 1086--1090, 2018.

\bibitem{Wang2018}
Hao Wang, Yitong Wang, Zheng Zhou, Xing Ji, Dihong Gong, Jingchao Zhou, Zhifeng
  Li, and Wei Liu,
\newblock ``Cosface: Large margin cosine loss for deep face recognition,''
\newblock {\em IEEE CVPR}, Jun 2018.

\bibitem{French1999}
Robert~M. French,
\newblock ``{Catastrophic forgetting in connectionist networks},''
\newblock in {\em Trends Cogn. Sci.}, 1999, vol.~3, pp. 128--135.

\bibitem{Goodfellow2013}
Ian~J. Goodfellow, Mehdi Mirza, Da~Xiao, Aaron Courville, and Yoshua Bengio,
\newblock ``{An empirical investigation of catastrophic forgetting in
  gradient-based neural networks},''
\newblock in {\em ICLR}, 2014.

\bibitem{Okabe2018}
Koji Okabe, Takafumi Koshinaka, and Koichi Shinoda,
\newblock ``{Attentive statistics pooling for deep speaker embedding},''
\newblock in {\em Interspeech}, 2018, vol. 2018-Septe, pp. 2252--2256.

\end{thebibliography}

\end{document}